\documentstyle[prl,aps]{revtex}
\begin{document}
\draft
\title{Basic limitations for entanglement catalysis}
\author{ZhengWei-Zhou\thanks{%
Email address: zwzhou@mail.ustc.edu.cn} and GuangCan-Guo\thanks{%
Email address: gcguo@ustc.edu.cn}}
\address{Laboratory of Quantum Communication and Quantum computation and \\
Department of Physics, University of Science and Technplogy of China,\\
Hefei 230026, P. R. China}
\maketitle

\begin{abstract}
In this paper we summarize the necessary condition for incomparable states
which can be catalyzed under entanglement-assisted LQCC (ELQCC). When we
apply an extended condition for entanglement transformation to
entanglement-assisted local manipulation we obtain a fundamental limit for
entanglement catalysts. Some relative questions are also discussed.
\end{abstract}




The existence of entanglement between spatially separated quantum systems is
at the heart of quantum information theory. In the light of recent progress
in quantum information theory entanglement is often viewed as the essential
resource for processing and transmitting quantum information and forms the
basis for many miraculous applications including quantum teleportation \cite
{bennett1}, quantum cryptography \cite{bennett2} and quantum communication 
\cite{cleve}. For many practically applications of this nonlocal resource,
one often restricts oneself to only preforming local quantum operations on
respective subsystems of entangled state and exchanging classical
information. In this paper transformations of this type will be referred to
as ``local transformation '', or `` LQCC '' for short. LQCC forms a
fundamental limit for quantification and manipulation of entanglement \cite
{bennett3,bennett4,bennett5}.

One can transform entangled quantum system from one situation to another
under LQCC. However, on average, the entanglement degree of the initial
system will be never enhanced. A quantitative way of expressing this fact is
in terms of so-called entanglement monotones ( EMs ) \cite{vidal1}. EMs
provide necessary restrictions for entanglement transformation under LQCC.
But, there arises naturally the question of what are the sufficient
conditions for entanglement transformation. One very fruitful approach was
achieved by Nielsen who provided a sufficient and necessary condition for
pure state entanglement transformations \cite{nielsen}. Inspired by
Nielsen's theory, Jonathan and Plenio investigated the behaviors of
entanglement-assisted local manipulation of pure quantum states and
presented a new concept called entanglement catalysis \cite{jonathan1}.
Afterwards, Eisert and Wilkens made further studies for catalysis of
entanglement manipulation for mixed states \cite{eisert}.

This remarkable phenomenon of entanglement catalysis can be described as
follows. Let $\rho _s$ and $\rho _t$ be source state and target state
respectively, which are taken from the state space S(H) over H, where H=$%
H_A\otimes H_B$ is the Hilbert space associated with a bipartite quantum
system consisting of part A and B. The target state $\rho _t$ can not be
reached by LQCC from the source state $\rho _s$ with certainty. But with the
assistance of a particular known entangled state $\rho _c$ taken from the
state space S($\widetilde{H}$) over $\widetilde{H}$ the transformation from $%
\rho _s\otimes \rho _c$ to $\rho _t\otimes \rho _c$ can be achieved by LQCC
with 100\% probability, where $\widetilde{H}$ is a tensor product $%
\widetilde{H}=\widetilde{H}_A\otimes \widetilde{H}_B$ of two Hilbert spaces
belonging to systems A and B respectively. The auxiliary entangled state $%
\rho _c$, which plays an indeed catalyst role in this process, is left in
exactly the same state and remain finally completely uncorrelated to the
quantum system of interest. This counter-intuitive effect demonstrate that
entanglement can be `` borrowed ''.

In this paper we first summarize the necessary condition for pure bipartite
incomparable states which can be catalyzed under entanglement-assisted LQCC
(ELQCC). Furthermore, starting from an extended condition for entanglement
transformation, we find that catalytic transition processes will provide a
fundamental limit for catalysts themselves.

Let us begin with Nielsen's theorem\cite{nielsen}.

Theorem ( Nielsen ): Let $\left| \Psi _1\right\rangle =\sum_{i=1}^n\sqrt{%
\alpha _i}\left| i_A\right\rangle \left| i_B\right\rangle $ and$\left| \Psi
_2\right\rangle =\sum_{i=1}^n\sqrt{\beta _i}\left| i_A\right\rangle \left|
i_B\right\rangle $ be pure bipartite states, where the Schmidt coefficients
are ordered according to $\alpha _1\geq \alpha _2\geq ...\geq \alpha _n>0$
and $\beta _1\geq \beta _2\geq ...\geq \beta _n>0$ respectively. ( We can
refer to such distributions as `` ordering Schmidt coefficients '', or OSCs.
) Then the transformation from $\left| \Psi _1\right\rangle $ to $\left|
\Psi _2\right\rangle $ with 100\% probability can be realized using LQCC iff
the OSCs\{$\alpha _i$\} are majorized by \{$\beta _i$\} \cite{marshall},
that is, iff for $1\leq l\leq n$ 
\begin{equation}
\sum_{i=1}^l\alpha _i\leq \sum_{i=1}^l\beta _i.  \label{eq1}
\end{equation}
Nielsen call the state $\left| \Psi _1\right\rangle $ and $\left| \Psi
_2\right\rangle $ incomparable if neither state can not convert into the
other with certainty. The current studies indicate only transitions between
incomparable states may be catalyzed\cite{jonathan1}. Then, under ELQCC what
conditions should be satisfied if the transformations between incomparable
states are possible? One naturally desires to find analogous Nielsen's
criterion. Unfortunately, we at present do not know what are sufficient
conditions for the existence of catalysts. To find appropriate catalysts one
has to resort to numerical search. Nevertheless, we may provide two
necessary conditions for entanglement catalysis.

Let $\left| \Psi _1\right\rangle =\sum_{i=1}^n\sqrt{\alpha _i}\left|
i_A\right\rangle \left| i_B\right\rangle $ and$\left| \Psi _2\right\rangle
=\sum_{i=1}^m\sqrt{\beta _i}\left| i_A\right\rangle \left| i_B\right\rangle $
( m$\leq $n ) be pure bipartite states with OSCs \{$\alpha _i$ ; i=1,
...,n\} \{$\beta _i$ , i=1, ... ,n\} ( here $\beta _i=0$ if i=m+1, ... , n
). Then $\left| \Psi _1\right\rangle $ can be converted into $\left| \Psi
_2\right\rangle $ with certainty under ELQCC only if 
\begin{equation}
(i)\text{ }\alpha _1\leq \beta _1\text{ , }\alpha _n\geq \beta _n
\label{eq2}
\end{equation}
\begin{equation}
(ii)\text{ }S\left( \rho _1\right) \geq S\left( \rho _2\right) ,  \label{eq3}
\end{equation}
where $S\left( \rho _i\right) $ is the marginal Von Neumann entropy of state 
$\left| \Psi _i\right\rangle .$ In ( ii ) saturation is reached iff $\left|
\Psi _1\right\rangle $ and $\left| \Psi _2\right\rangle $ are locally
unitarilly equivalent.

Condition ( i ) has been proved in \cite{jonathan1}. For condition ( ii ) we
offer a brief proof in the following.

Proof: Suppose that $\left| \Psi _1\right\rangle $ can be transformed into $%
\left| \Psi _2\right\rangle $ under ELQCC. Then there exists a catalyst $%
\left| \Psi _c\right\rangle $ such that $\left| \Psi _1\right\rangle \left|
\Psi _c\right\rangle $ can be converted to $\left| \Psi _2\right\rangle
\left| \Psi _c\right\rangle $ with 100\% probability under LQCC. ( For
simplicity, in the next context we indicate this process by $\left| \Psi
_1\right\rangle \left| \Psi _c\right\rangle \rightarrow \left| \Psi
_2\right\rangle \left| \Psi _c\right\rangle .$ ) In light of the
non-increase of partial entropy under LQCC\cite{bennett3} and additivity of
entropy\cite{wehrl} the inequality (\ref{eq3}) can be obtained. In order to
prove the saturation case we may refer to Theorem 1 in \cite{bennett6}. If $%
\left| \Psi _1\right\rangle $ and $\left| \Psi _2\right\rangle $ are two
marginally isentropic pure states, they are either locally unitarilly
equivalent or else LQCC-incomparable. A manifest fact is that $\left| \Psi
_1\right\rangle \left| \Psi _c\right\rangle $ and $\left| \Psi
_2\right\rangle \left| \Psi _c\right\rangle $ is locally unitarilly
equivalent if they are marginal isentropic. Furthermore, we may deduce that $%
\left| \Psi _1\right\rangle $ and $\left| \Psi _2\right\rangle $ are locally
unitarilly equivalent.

The above conditions just put forward limits for the relationship between
source and target. Suppose that two incomparable states to hold inequality (%
\ref{eq2}) and (\ref{eq3}) can be ``catalyzed''. Are there any limitations
for potential catalysts? To obtain our theorem the following lemma is
necessary.

Lemma 1: Let $\left| \Psi _p\right\rangle $ be p$\times $p-level maximal
entangled state. Then the maximal probability of obtaining state $\left|
\Psi _2\right\rangle $ from $\left| \Psi _1\right\rangle $ by means of LQCC
is just equal to that of $\left| \Psi _1\right\rangle \left| \Psi
_p\right\rangle \rightarrow \left| \Psi _2\right\rangle \left| \Psi
_p\right\rangle $, i.e. 
\begin{equation}
P_{\max }\left( \left| \Psi _1\right\rangle \rightarrow \left| \Psi
_2\right\rangle \right) =P_{\max }\left( \left| \Psi _1\right\rangle \left|
\Psi _p\right\rangle \rightarrow \left| \Psi _2\right\rangle \left| \Psi
_p\right\rangle \right) .  \label{eq4}
\end{equation}

Proof: The maximal probability of $\left| \Psi _1\right\rangle \rightarrow
\left| \Psi _2\right\rangle $ has been achieved by Vidal\cite{vidal2}: 
\begin{equation}
P_{\max }\left( \left| \Psi _1\right\rangle \rightarrow \left| \Psi
_2\right\rangle \right) =\min_{l\in \left[ 1,n\right] }\frac{%
\sum_{i=l}^n\alpha _i}{\sum_{i=l}^n\beta _i}=\min_{l\in \left[ 1,n\right] }%
\frac{E_l\left( \left| \Psi _1\right\rangle \right) }{E_l\left( \left| \Psi
_2\right\rangle \right) },  \label{eq5}
\end{equation}
where $E_l\left( \left| \Psi _1\right\rangle \right) =\sum_{i=l}^n\alpha _i$
is called entanglement monotone of state $\left| \Psi _1\right\rangle $. $%
\left| \Psi _1\right\rangle \left| \Psi _p\right\rangle $ and $\left| \Psi
_2\right\rangle \left| \Psi _p\right\rangle $ have separately p-fold
degenerate OSCs \{$\alpha _i^{\prime }$ ; i=1, ... ,pn\} and \{$\beta
_i^{\prime }$ ; i=1, ... ,pn\}. if $l$ is a number between $pk+1$ and $%
p\left( k+1\right) $, $k=0,1,...,n-1$, we have: 
\begin{equation}
\frac{E_l\left( \left| \Psi _1\right\rangle \left| \Psi _p\right\rangle
\right) }{E_l\left( \left| \Psi _2\right\rangle \left| \Psi _p\right\rangle
\right) }=\frac{\frac{l-pk-1}pE_{k+2}\left( \left| \Psi _1\right\rangle
\right) +\frac{p(k+1)-l+1}pE_{k+1}\left( \left| \Psi _1\right\rangle \right) 
}{\frac{l-pk-1}pE_{k+2}\left( \left| \Psi _2\right\rangle \right) +\frac{%
p(k+1)-l+1}pE_{k+1}\left( \left| \Psi _2\right\rangle \right) },  \label{eq6}
\end{equation}
where $E_{n+1}\left( \left| \Psi _1\right\rangle \right) =E_{n+1}\left(
\left| \Psi _2\right\rangle \right) =0.$ Without loss of generality, let us
set $\frac{E_{k+2}\left( \left| \Psi _1\right\rangle \right) }{E_{k+2}\left(
\left| \Psi _2\right\rangle \right) }>\frac{E_{k+1}\left( \left| \Psi
_1\right\rangle \right) }{E_{k+1}\left( \left| \Psi _2\right\rangle \right) }
$. Taking advantage of the equivalence: 
\begin{equation}
\frac ab<\frac{a+c}{b+d}\Leftrightarrow \frac ab<\frac cd,  \label{eq7}
\end{equation}
we have the following relation: 
\begin{equation}
\frac{E_l\left( \left| \Psi _1\right\rangle \left| \Psi _p\right\rangle
\right) }{E_l\left( \left| \Psi _2\right\rangle \left| \Psi _p\right\rangle
\right) }\geq \frac{E_{k+1}\left( \left| \Psi _1\right\rangle \right) }{%
E_{k+1}\left( \left| \Psi _2\right\rangle \right) }.  \label{eq8}
\end{equation}
Here, the saturation holds iff $l=pk+1$. Therefore, 
\[
P_{\max }\left( \left| \Psi _1\right\rangle \left| \Psi _p\right\rangle
\rightarrow \left| \Psi _2\right\rangle \left| \Psi _p\right\rangle \right)
=\min_{l\in \left[ 1,pn\right] }\frac{\sum_{i=l}^{pn}\alpha _i^{\prime }}{%
\sum_{i=l}^{pn}\beta _i^{\prime }}
\]
\begin{equation}
=\min_{l\in \left[ 1,n\right] }\frac{\sum_{i=l}^n\alpha _i}{%
\sum_{i=l}^n\beta _i}=P_{\max }\left( \left| \Psi _1\right\rangle
\rightarrow \left| \Psi _2\right\rangle \right) .  \label{eq9}
\end{equation}
This completes the proof of lemma 1.

The following theorem provide us with a fundamental limit for catalysts.

Theorem 1: Successful transformation from $\left| \Psi _1\right\rangle $ to $%
\left| \Psi _2\right\rangle $ can be reached under ELQCC only if any of a p$%
\times $p-level catalyst with OSCs \{$\gamma _i$ ; i=1, ... , p\} meet the
following relation: 
\begin{equation}
p\gamma _p\leq P_{\max }\left( \left| \Psi _1\right\rangle \rightarrow
\left| \Psi _2\right\rangle \right) .  \label{eq10}
\end{equation}

Proof: Based on lemma 1 we have 
\begin{eqnarray}
P_{\max }(\left| \Psi _1\right\rangle  &\rightarrow &\left| \Psi
_2\right\rangle )=P_{\max }(\left| \Psi _1\right\rangle \left| \Psi
_p\right\rangle \rightarrow \left| \Psi _2\right\rangle \left| \Psi
_p\right\rangle )  \nonumber \\
&\geq &P_{\max }(\left| \Psi _1\right\rangle \left| \Psi _c\right\rangle
\rightarrow \left| \Psi _2\right\rangle \left| \Psi _p\right\rangle ) 
\nonumber \\
&\geq &P_{\max }(\left| \Psi _2\right\rangle \left| \Psi _c\right\rangle
\rightarrow \left| \Psi _2\right\rangle \left| \Psi _p\right\rangle )
\label{eq11} \\
&\geq &P_{\max }(\left| \Psi _c\right\rangle \rightarrow \left| \Psi
_p\right\rangle ),  \nonumber
\end{eqnarray}
where $\left| \Psi _p\right\rangle $ and $\left| \Psi _c\right\rangle $ are p%
$\times $p-level maximal entangled state and catalyst state respectively.
The first inequality is satisfied due to the following fact: one can look on
the process $\left| \Psi _1\right\rangle \left| \Psi _p\right\rangle
\rightarrow \left| \Psi _1\right\rangle \left| \Psi _c\right\rangle $ as one
of steps in the transformation from $\left| \Psi _1\right\rangle \left| \Psi
_p\right\rangle $ to $\left| \Psi _2\right\rangle \left| \Psi
_p\right\rangle $ while the transformation from $\left| \Psi _p\right\rangle 
$ to $\left| \Psi _c\right\rangle $ is deterministic under LQCC. If any an
intermediate state can be deterministically arrived the probability from
this intermediate state to target state can not be higher than the maximal
probability from source to target. Otherwise, this will lead to
contradiction. Depended on the similar options the second inequality can
also be obtained. While, the last inequality is obvious. On the basis of (%
\ref{eq5}) and (\ref{eq7}), the maximal probability of $\left| \Psi
_c\right\rangle \rightarrow \left| \Psi _p\right\rangle $ is as follows: 
\begin{equation}
P_{\max }(\left| \Psi _c\right\rangle \rightarrow \left| \Psi
_p\right\rangle )=p\gamma _p.  \label{eq12}
\end{equation}
We thus complete the proof of our theorem.

As a direct consequence, this theorem can supply some concrete limits for
entanglement catalysts under some special circumstances. For example\cite
{jensen}, our choice of Schmidt coefficients for source state$\left| \Psi
_1\right\rangle $ and target state$\left| \Psi _2\right\rangle $ is $\alpha
_1=\alpha _2=0.31$, $\alpha _3=0.3$, $\alpha _4=\alpha _5=0.04$, $\beta
_1=0.48$, $\beta _2=0.24$, $\beta _3=\beta _4=0.14$, $\beta _5=0$. Taking
advantage of this theorem we can easily conclude that 2$\times $2-level
catalysts do not exist for the process $\left| \Psi _1\right\rangle
\rightarrow \left| \Psi _2\right\rangle $. A detailed analysis is as
follows. Assume that there exists a 2$\times $2-level catalyst $\left| \Psi
_c\right\rangle $, with OSCs \{ $x,1-x$ \}. In view of the inequality
relation of (\ref{eq10}) we have $2\times \left( 1-x\right) \leq \frac 47%
,\Rightarrow x\geq \frac 57$. Hence, the first three OSCs of $\left| \Psi
_1\right\rangle \left| \Psi _c\right\rangle $ must be $0.31x$, $0.31x$, $0.3x
$. Similarly, the first three OSCs of $\left| \Psi _2\right\rangle \left|
\Psi _c\right\rangle $ is either $0.48x$, $0.24x$, $0.48(1-x)$ or $0.48x$, $%
0.24x$, $0.14x$. No matter which cases take place the relation $%
\sum_{i=1}^3\alpha _i^{\prime }>\sum_{i=1}^3\beta _i^{\prime }$ must be
satisfied. ( Here, \{$\alpha _i^{\prime }$\} and \{$\beta _i^{\prime }$\}
refer to OSCs of $\left| \Psi _1\right\rangle \left| \Psi _c\right\rangle $
and $\left| \Psi _2\right\rangle \left| \Psi _c\right\rangle $
respectively.) We thus have $1-\sum_{i=1}^3\alpha _i^{\prime }=E_4\left(
\left| \Psi _1\right\rangle \left| \Psi _c\right\rangle \right)
<1-\sum_{i=1}^3\beta _i^{\prime }=E_4\left( \left| \Psi _2\right\rangle
\left| \Psi _c\right\rangle \right) $. In view of Nielsen's theorem this
implies there exist no $2\times 2$-level entangled states to hold that \{$%
\alpha _i^{\prime }$\} can be majorized by \{$\beta _i^{\prime }$\}. Seeing
reference \cite{jensen}, we know $\left| \Psi _2\right\rangle $ itself just
a catalyst of the process $\left| \Psi _1\right\rangle \rightarrow \left|
\Psi _2\right\rangle $. From this example we find that higher dimensional
entangled states have exactly more powerful capability of catalysis than
lower dimensional entangled states. The reasons rely on this fact: theorem 1
provides only quite slack bounds for the structure of OSCs of catalysts when
marginal Hilbert space of entanglement catalysts have high dimensions. In
other words, more degrees of freedom conceals in higher dimensional
entanglement. When applying our theorem to generalized cases, we may acquire
some interesting corollaries.

According to the inequality (\ref{eq10}), entanglement catalysts can be
divided into two sorts. We call those catalysts saturating inequality (\ref
{eq10}) ``saturated catalysts'', or else ``non-saturated catalysts''. At
present, we do not know whether there are saturated catalysts for all pairs
of convertible incomparable states under ELQCC. However, any of a pair of
incomparable states with saturated catalysts must meet the following
corollary.

Corollary 1: $\left| \Psi _1\right\rangle $ can be deterministically
transformed into $\left| \Psi _2\right\rangle $ by using saturated
catalysts-assisted LQCC only if 
\begin{equation}
P_{\max }(\left| \Psi _1\right\rangle ^{\otimes n}\rightarrow \left| \Psi
_2\right\rangle ^{\otimes n})\geq P_{\max }(\left| \Psi _1\right\rangle
\rightarrow \left| \Psi _2\right\rangle ).  \label{eq13}
\end{equation}

Proof: Similar to the proof of theorem 1, we have 
\[
P_{\max }(\left| \Psi _1\right\rangle ^{\otimes n}\rightarrow \left| \Psi
_2\right\rangle ^{\otimes n})=P_{\max }(\left| \Psi _1\right\rangle
^{\otimes n}\left| \Psi _p\right\rangle \rightarrow \left| \Psi
_2\right\rangle ^{\otimes n}\left| \Psi _p\right\rangle ) 
\]
\[
\geq P_{\max }(\left| \Psi _1\right\rangle ^{\otimes n}\left| \Psi
_c\right\rangle \rightarrow \left| \Psi _2\right\rangle ^{\otimes n}\left|
\Psi _p\right\rangle )\geq P_{\max }(\left| \Psi _2\right\rangle ^{\otimes
n}\left| \Psi _c\right\rangle \rightarrow \left| \Psi _2\right\rangle
^{\otimes n}\left| \Psi _p\right\rangle ) 
\]
\begin{equation}
\geq P_{\max }(\left| \Psi _c\right\rangle \rightarrow \left| \Psi
_p\right\rangle )=p\gamma _p=P_{\max }(\left| \Psi _1\right\rangle
\rightarrow \left| \Psi _2\right\rangle ).  \label{eq14}
\end{equation}
This proves corollary 1.

One can make further extensions for the concept of entanglement-assisted
transformations. For instance, there is a case of ``quasi-catalysis'' in
which the transformation from $\left| \Psi _1\right\rangle $ to $\left| \Psi
_2\right\rangle $ can not be preformed with certainty even under ELQCC, but
the optimal probability of transformation may still be increased\cite
{jonathan1}. Based on the analogous analysis, we have:

Corollary 2: The probability of transition from state $\left| \Psi
_1\right\rangle $ to $\left| \Psi _2\right\rangle $ can be enhanced to $%
P^{\prime }$ with the assistance of a $p\times p$-level entangled state $%
\left| \Psi _c\right\rangle =\sum_{i=1}^p\sqrt{\gamma _i}\left|
i\right\rangle \left| i\right\rangle $ only if 
\begin{equation}
p\gamma _p\leq P_{\max }(\left| \Psi _1\right\rangle \rightarrow \left| \Psi
_2\right\rangle )/P^{\prime }.  \label{eq15}
\end{equation}

Of late, one begins to consider practical applications for entanglement
catalysis. Two schemes for quantum secure identification using catalysts
have been presented by Barnum \cite{barnum} and Jensen et.al \cite{jensen}
respectively. These proposals using entanglement catalysts have an
attracting prospect relying on this fact: entanglement catalysts will not be
depleted during quantum information processes, i.e. a protracted
characteristic entanglement between quantum users may be employed
repeatedly. However, Barnum's protocol has been shown to be insecure \cite
{jensen}. On the condition that all quantum operations are error-free and
that the quantum channel is noiseless the quantum authentication protocol
presented by Jensen and Schark appears to be secure even in the presence of
an eavesdropper who has complete control over both classical and quantum
communication channels at all times\cite{jensen}. How to develop secure and
unjammable quantum authentication schemes is attracting more and more
attention.

In conclusion, we have shown some necessary limitations for pure bipartite
incomparable states which can be catalyzed under ELQCC and catalysts
themselves. We find that the product of dimension and the final OSC of
catalysts restricts the ability of catalysts. We believe that the results of
the present paper can help in deeper understanding of entanglement-assisted
local manipulations.

\vskip 5mm

This project was supported by the National Nature Science Foundation of
China and the Doctoral Education Fund of the State Education Commission of
China.

\end{document}